\documentclass[twocolumn,superscriptaddress,prl,rmp]{revtex4-1}
\usepackage{amsmath}
\usepackage{amssymb}

\usepackage{xcolor}
\usepackage{graphicx}% Include figure files
\usepackage{dcolumn}% Align table columns on decimal pointd
\usepackage{bm}% bold math
\usepackage{verbatim}
\usepackage{float}
\floatstyle{plaintop}
\restylefloat{table}

\usepackage{mathtools}
\usepackage{commath}

\usepackage{subcaption}

\usepackage{hyperref} %ALWAYS BEFORE CLEVREF
\usepackage[noabbrev,nameinlink]{cleveref}
\creflabelformat{equation}{#2\textup{#1}#3}

\usepackage{relsize}
\usepackage[colorinlistoftodos]{todonotes}

\usepackage{fixltx2e}
\usepackage{xspace}
\newcommand{\nox}{\texorpdfstring{NO\textsubscript{x}}{NOx}\xspace}
\newcommand{\nhthree}{\texorpdfstring{NH\textsubscript{3}}{NO3}\xspace}
\newcommand{\co}{\texorpdfstring{CO\textsubscript{2}}{CO2}\xspace}
\newcommand{\otwo}{\texorpdfstring{O\textsubscript{2}}{O2}\xspace}
\newcommand{\othree}{\texorpdfstring{O\textsubscript{3}}{O3}\xspace}

\newcommand{\cb}{\texorpdfstring{CAN-bus}{cb}\xspace}
\newcommand{\tslm}{2000}
\newcommand{\tslkm}{2}

\usepackage{siunitx}
\newcommand*{\imagePathTrue}{output_old_values}

\DeclareSIUnit\kmh{km/h}
\usepackage{natbib}
\setcitestyle{authoryear,square, comma, numbers,sort&compress, super}
\usepackage{etoolbox}
\makeatletter
\patchcmd\linenumberpar{\@LN@parpgbrk}{\penalty\@LN@parpgpen\relax}{}{}
\makeatother

\begin{document}

%\preprint{APS/123-QED}
\title{Impact of Signalized Intersections on \co and \nox Emissions of Heavy Duty Vehicles}% Force line breaks with \\
%\title{Impact of Signalized Intersections in Fuel Consumption and \nox Emissions in Heavy Duty Vehicles}% Force line breaks with \\
\author{Nicolás Deschle}
\email{nico.deschle@tno.nl}
\affiliation{TNO, Sustainable Urban Mobility and Safety group, The Hague, The Netherlands}
\author{Ernst Jan van Ark}
\affiliation{TNO, Sustainable Urban Mobility and Safety group, The Hague, The Netherlands}
\author{René van Gijlswijk}
\affiliation{TNO, Sustainable Transport and Logistics group, The Hague, The Netherlands}
\author{Robbert Janssen}
\affiliation{TNO, Sustainable Transport and Logistics group, The Hague, The Netherlands}

\date{\today} 
\begin{abstract}
Pollutant emissions have been a topic of interest in the last decades. Not only environmentalists but also governments are taking rapid action to reduce emissions. As one of the main contributors, the transport sector is being subjected to strict scrutiny to ensure it complies with the short and long-term regulations. The measures imposed by the governments clearly involve, all the stakeholders in the logistics sector, from road authorities and logistic operators to truck manufacturers.
Improvement of traffic conditions is one of the perspectives in which the reduction of emissions is being addressed. Optimization of traffic flow, avoidance of unnecessary stops, control of the cruise speed, and coordination of trips in an energy-efficient way are necessary steps to remain compliant with the upcoming regulations. In this study, we have measured the \co and \nox emissions in heavy-duty vehicles while traversing signalized intersections and we examined the differences between various scenarios. We found that avoiding a stop can reduce \co and \nox emissions on \SI{0.32}{\kg} and \SI{1.8}{\g}, respectively. These results put traffic control in the main scene as a yet unexplored dimension to control pollutant emissions, enabling the authorities to more accurately estimate cost-benefit plans for traffic control system investments.

%We found that in good traffic conditions avoiding a stop translates to a $0.12\ l$ of fuel saved and a reduction of $0.18\ g$ of \nox.

\end{abstract}
\maketitle
\onecolumngrid
%\appendix

%\textcolor{red}{target journals:\\
%easier would be (current format): https://www.journals.elsevier.com/transportation-research-part-d-transport-and-environment 4.051\\
%best would be: https://www.journals.elsevier.com/science-of-the-total-environment 6.55\\
%other best: https://iopscience.iop.org/journal/1748-9326 impact factor 6.192\\
%another option: https://www.journals.elsevier.com/atmospheric-environment 4.012\\
%another last option: https://journals.sagepub.com/home/pid\\
%https://www.journals.elsevier.com/transportation-research-procedia\\}
%\textcolor{red}{check this link for the guidelines (TR part D). \url{https://www.elsevier.com/journals/transportation-research-part-d-transport-and-environment/1361-9209/guide-for-authors}}

\section{Introduction}
%\textcolor{red}{State the objectives of the work and provide an adequate background, avoiding a detailed literature survey or a summary of the results.}\\

Despite tremendous improvements in recent years, world air quality is still far from the levels that do not represent a risk for human health and the environment. Nitrogen oxides (\nox), together with particulate matter, ground level ozone (\othree), and ammonia (\nhthree) are among the most problematic air pollutants \cite{mendozavillafuerte2017nox,liu2020identifying,abera2020air,cohen2017estimates,burnett2018global}. At the same time, the concentration of greenhouse gases such as \co keeps increasing in the atmosphere. Road transportation is not only one of the main causes of \nox and \co emissions, but it is also the main factor in the logistic costs. Furthermore, a high percentage of this emissions occurs in specific corridors where most of the freight traffic takes place. Reducing the emissions and fuel usage in these specific corridors can have a huge economic and environmental impact.

\subsection{Fuel Consumption, \co and \nox Emissions}
Depending on the degree of development of the country, domestic logistics costs account for $5 \%$ to $20 \%$ of a country's GDP of which about $60 \%$ are transportation costs \citep{ittmann2010state,havenga2012quantifying}. %The main transportation means are rail, road and to a lesser degree inland waterways.
Although the global road to rail modal split ratio is estimated to be 60:40, this varies significantly from one country to another. Specifically, in Latin America and China, the percentage of rail transport is below $25 \%$ and it is under $40 \%$ for the United States~\cite{kaack2018decarbonizing}. The European Union is more heterogeneous; the estimations of the road-rail modal split vary from $34 \%$ \cite{kaack2018decarbonizing} to $17 \%$ \citep{blauwens2007towards}. Rail dominates in a few countries that share certain characteristics, i.e., they span large areas with very irregular population distribution as Australia, Russia and Canada. Furthermore, in the last ten years, the share of road freight activity showed to be increasing with respect to rail freight worldwide \citep{kaack2018decarbonizing}.

At the same time, the transport industry is a significant source of \nox%, \pms
emission as well as greenhouse gases such as \co. As of 2015, the transportation sector was responsible for $7 \%$ of the total \co energy-related emissions \citep{kaack2018decarbonizing}. 
In the European Union, transportation overall is responsible for $14 \%$ of \co emissions as well as a main contributor to the overall \nox emissions \citep{dekker2012operations}.
Road and rail transport emissions together account for $32 \%$ %and $21 \%$ 
of the total \nox %and \pmtwo
of the United States \citep{hwang2013freight,bickford2012emissions}%.
.%, respectively.
Furthermore, freight transport, which is highly dominated by trucks, constitutes $25 \%$ of U.S. total \co emissions \citep{hwang2013freight,ang2005assessing}. Altogether, these figures call for urgent action in road transportation in order to let the logistics expand allowing for an economic growth with smaller impact on air quality as is described in the emissions goals in particular in the European Union.

The last two decades have already shown a shift in freight transport research towards sustainability and safety affecting heavy-duty transport vehicles in a number of ways. Different options have been and are being explored at different levels ranging from exploring different energy sources to automation and coordination. The changes in the transport of goods could be classified as follows; vehicle energy consumption and efficiency, control of emissions, and finally automation and coordination, both between vehicles and with the network infrastructure.

\subsection{Transport corridors}
Special attention must be given to the main corridors where a high percentage of heavy-duty transport takes place. These corridors are also known for playing a critical role in the expansion economy \citep{roberts2019transport}. Roberts and coworkers found that although the economic benefits are significant, large transport infrastructure projects can detrimentally impact the environment. Clearly, focusing on these specific routes is key to have a high impact on overall transport. It is in this context that in 2013, the EU started the development of the Trans-European Transport Network (TEN-T), which aims to improve the European transport network focused to allow sustainable growth of the economy with special attention on the identified nine \emph{Core Network Corridors} \citep{weenen2016study,eu2013corridors,tent2013,oeberg2017major} of which $39\%$ of the road corridors are off highways \cite{cedr2017tent}. Projects also at national level support these initiatives, as are the cases of the \emph{top corridors} program and the \emph{Connected Transport Corridors} consortium~\cite{mirt201xtopcorridors,topcorridors} % \textcolor{red}{more citations!}
in the Netherlands which aim to scale up the currently available technology in four main Dutch corridors in order to achieve a more efficient, sustainable and safety transport of goods. In particular, one of the objectives is the implementation and analysis of coordination systems between the vehicles and the infrastructure. 

%Is in this context that the Ministry and the local Dutch governments together with private investors are carrying out a national deployment of smart mobility applications focused in four connected transport corridors. This consortium aims to scale the current available technology on these four corridors in order to achieve a more efficient, sustainable and safety transport of goods. In particular, one of the objectives is the implementation and analysis of coordination systems between the vehicles and the infrastructure.

\subsection{Emissions and Traffic Control}
%\subsection{Traffic Control Systems}
%\subsubsection{Cooperative Traffic Lights}
Reducing the number of stops of vehicles can lead to a drop in travel time, fuel consumption and undesired emissions (\co, \nox) at the same time that increases the safety on road (see \cite{peters2009reducing} and the references therein, for example \cite{barth2008real}). Although initially aimed to reduce travel time rather than increase sustainability, methods to maintain a steady flow in main roads have been around since the implementation of the green waves studied since the late 1910s \cite{tully1976synthesis,guberinic2008optimal}. Further steps in this direction were green waves influenced by induction loops where the fixed or manually adjustable schedule was replaced by a dynamic algorithm with the traffic flow on specific points as input. The application of induction loops for traffic control was introduced in the 1960s \cite{klein2006traffic} and has been under development since then. Some of the currently widely used traffic control systems based on induction loops are SCOOT \cite{robertson1991optimizing} and SCAT \cite{sims1980sydney}. In the Netherlands the VECOM and VETAG systems have been in use since the 1970s \cite{vetag,middleham1976vetag,meyer1975points,ros1986vecom}.
%\cite{roess2011traffic} %These unidirectional communication systems are able to identify vehicles and influence the traffic light network. 
Since then, diverse efforts have been put in developing communication systems between vehicles, infrastructure and traffic management centers. This technology is now widely referred to as \emph{intelligent transport systems} (ITS). PROMETHEUS, probably the first European research program on ITS, dates back to the 1980s \cite{festag2014cooperative}. It was around the same time that the Interstate Surface Transportation Efficiency Act (ISTEA) in the United States issued the theme of ITS \cite[p.~625]{roess2011traffic}.
In the early 2000s, the ubiquitous availability of Global Positioning System (GPS) devices and wireless communication boosted ITS technology developing \emph{Cooperative}-ITS (C-ITS) systems among which intelligent traffic light control systems (iTLC) can be found. iTLCs not only receive information about traffic in advance both from induction loops and from cell-phone applications allowing to coordinate the traffic reducing the number of stops, but these systems also communicate bidirectionally sending back information to the vehicle or the driver, being able to concede priority and determine the optimal speed based on real-time data. iTLC systems come with a great opportunity since they can contribute to achieving the environmental goals by optimizing traffic such that emissions are minimized. %This is one of the improvements being implemented on the CTC also in the context of the Dutch consortium Talking Traffic.

\subsection{Previous studies}
Several projects in the last decade addressed the impact of the interaction between traffic control and behavior on pollutant emissions. Despite most of these studies aim to bridge the gap between laboratory tests (as those done on a dynamometer) and real-world emissions, very few of them involve real-world field measurements. Much research is focused on theoretical optimization such determining optimal speed in terms of \co and \nox \cite{chang2005vehicle,wang2017fuel} or under the relation between emissions on different traffic conditions based on level or urbanization. In 2017, Wang and Reakha developed a model to determine real-world optimal accelerations and cruise speed for heavy duty vehicles \cite{wang2017fuel}. Guardiola and co-workers studied the \nox emissions on passenger cars when approaching a traffic light using three modelled speed profiles measured on a chassis dynamometer \cite{guardiola2019potential} finding savings in the range of $7.5-12\ \%$ and $13-32\ \%$ for fuel emissions and \nox, respectively. Real world measurements were done by Meneguzzer et. al comparing \nox and \co emissions on passenger cars on a route were traffic lights had been replaced by a roundabout. Although not statistically significant, their results for \co supports our research~\cite{meneguzzer2017comparisson}.
Several European projects measured fuel consumption differences under traffic control measures. The Freilot project, which is probably the largest project to date with measurements on 177 trucks over 12 months around Europe, reported savings up to $13\ \%$ in fuel consumption on HDV by means of energy efficient intersection control \cite{freilot2012} followed by Compass4D, where 45 trucks with C-ITS capabilities (priority and speed advice) where measured found fuel savings in the order of $20-60\ g$ per $km$ \cite{compass4d2015}.
Despite the vast evidence supporting the potential of traffic control systems to reduce emissions, the heterogeneity of the methods and relatively limited properly published results make the comparison and scale up of these results very involved. An accurate and clean figure obtained from real world measurements is still missing and helping to fill that gap is the goal of this paper.
%Despite all providing positive results in terms of fuel consumption the observables and the methodology employed on them differ from each other. Freilot, connected transport experience week, C-the difference project and the compass4D designs had both traffic signal priority and advisory optimal speed on board but only in some of the measurements. Moreover, regarding the main outcome measure, the fuel consumption, it was not always measured directly. In the Freilot project some vehicles measured the fuel consumption from the truck (\cb), while others computed it indirectly from GPS and road network data [4 p 31, figure 22]. 
%Research so far shows conclusively positive impact of the C-ITS systems on fuel consumption on heavy duty trucks. The different methods employed on the among the projects together with different methodologies and experimental conditions make the results from different studies hard to compare. An accurate and precise measurement of fuel consumption change on the presence of intelligent traffic lights, key to asses economic impact and cost-benefit analysis, remains missing.
%\begin{table}[H]
%\centering
%\begin{tabular}{l|l|l|l}
%Date & Study Name & Description & Main results \\ \hline
%\multicolumn{1}{c|}{double} & \multicolumn{1}{c|}{a\_ref} & Requested acceleration & $m/s^2$ \\ \hline
%\end{tabular}
%\caption{Overview of different studies carried out regarding the impact of C-ITS on fuel consumption and pollutant emissions.}
%\end{table}
\section{Material and methods}
\label{sec:methods}
%\textcolor{red}{Provide sufficient details to allow the work to be reproduced by an independent researcher. Methods that are already published should be summarized and indicated by a reference. If quoting directly from a previously published method, use quotation marks and also cite the source. Any modifications to existing methods should also be described.}
The current study employed physical measurements of emissions, fuel consumption, speed and position of a group of vehicles. After identifying the location of the traffic lights, the data around the crossing was selected to study the different speed and emission profiles and in particular, the differences in emissions of different approaches. Furthermore, the data were sibjected to a process of filtering, analysis and enrichment. The following subsections are dedicated to explaining; first, the physical measurements and the properties of the vehicles employed; second, the pre-processing and enrichment of the data acquired and third, the analysis which includes filtering and reference to the statistical methods used.
\subsection{Physical measurements and data acquisition}
%\begin{figure}[t]
%\centering
%\includegraphics[height=150 pt]{\imagePathTrue/sems_welding}
%\caption{Fuel consumption for the three scenarios.}
%\label{fig:sems_welding}
%\end{figure}
%
%\begin{figure}[t]
%\raggedleft
%\includegraphics[width=\columnwidth]{\imagePathTrue/sems_box}
%\caption{Fuel consumption for the three scenarios.}
%\label{fig:sems_welding}
%\end{figure}
The measurements have been done on five Euro VI DAF trucks with engines with %rated
power between $315$ and $355\ \si{\kW}$ (see table \ref{tab:truckSpecs} for details). These data was acquired in the context of the integrator project \cite{integrator2019} and was made available for this study. The total weight of the vehicles was also estimated by means of the truck features and their kinematics over all the datasets. The mean and standard deviation of the total mass (tractor + load) of the vehicles including all the trips were $25.75\ \si{\tonne}$ and $9.05\ \si{\tonne}$
%$38.16\ \si{\tonne}$ and $14.55\ \si{\tonne}$
, respectively. This was estimated for each trip as described in the Appendix. The data were acquired between the months of September 2019 and February 2020 while the trucks realized their normal operations % (naturalistic behavior)
in The Netherlands region, traversing a total of \SI{230000}{\km}. Although the routes included segments outside The Netherlands, only the segments done on the Dutch Network were used in this analysis.
%\begin{table}[H]
%\centering
%\begin{tabular}{c|c|c|c}
%Vehicle ID & Engine & Power [\si{\kW}] & Plate \\ \hline \hline
%3 & XF 440 FTG & 320 & 18BFT5 \\ 
%4 & XF 460 FTG & 341 & 58BJN6 \\ 
%1 & XF 440 FT & 324 & 77BGH7 \\ 
%2 & XF 480 FTG & 353 & 88BLD6\\ 
%5 & XF 480 MX-13 & 355 & 3100KVF (spain) \\ 
%\end{tabular}
%\caption{\textcolor{red}{remove plates and ID}Details of all the vehicles included in the study. They were DAF trucks with similar characteristics. The rated power was between 320 and 355 \si{KW}.}
%\label{tab:truckSpecs} 
%\end{table}
\begin{table}[H]
\centering
\begin{tabular}{c||c}
Engine & Power [\si{\kW}] \\ 
\hline
\hline
XF 440 FTG & 320 \\ 
XF 460 FTG & 341 \\ 
XF 440 FT & 324 \\
XF 480 FTG & 353 \\ 
XF 480 MX-13 & 355 \\
\end{tabular}
\caption{Details of all the vehicles included in the study. They were DAF trucks with similar characteristics. The rated power was between 320 and 355 \si{KW}.}
\label{tab:truckSpecs} 
\end{table}

\begin{figure}[H]
%\raggedleft
\centering
\includegraphics[width=.4\columnwidth]{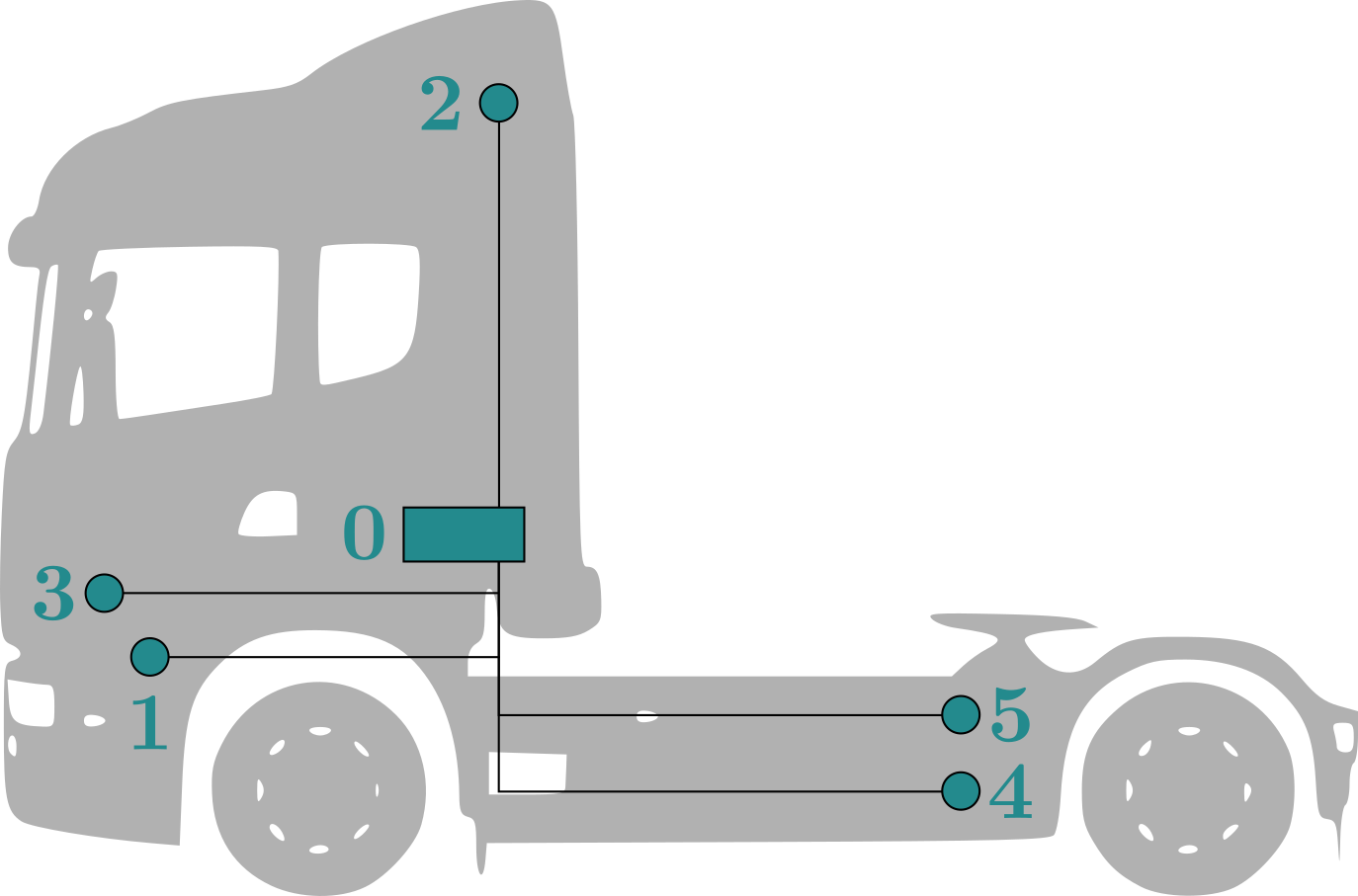}
\caption{Diagram of a tractor with the location of the different measuring systems installed on the vehicle depicted in green. (1) \cb sensor, (2) GPS antenna, (3) Air pressure sensor, (4) Temperature sensor and (5) Exhaust sensors of \nox and \otwo. (0) Indicates the location of the onboard unit system responsible for logging the data.}
\label{fig:truck}
\end{figure}

\subsubsection{SEMS system}
Onboard measurements have been done by means of the \emph{Smart Emissions Measurement System} (SEMS) developed by TNO. This system provides a simple and easy-to-use yet robust and reliable emissions monitoring solution. It has been tested extensibly since its development in 2012 (see for example, \cite{kadijk2015road,vermeulen2012smart,kadijk2017nox}) and there is an ongoing project to develop its industrialization and large-scale deployment \cite{heepen2019sems}.
A series of measurement systems need to be installed on the vehicles, whereas some data have been obtained from devices already installed by the original manufacturer.

The first group encompasses a GPS receiver as well as sensors to measure \nox concentrations, \otwo concentrations, \nhthree concentrations, and exhaust gas temperature. The sensors were mounted near the end of the exhaust line, in threaded bushes welded through the wall of the tailpipe. Regarding the second group, data from the \cb was acquired, such as velocity, throttle position and rotational speed of the engine. All data is sampled once per second and gathered in a data logger located in the cabin, as depicted in Figure \ref{fig:truck}. This small computer was set up to start (stop) acquisition when the ignition was turned on (off).
A built-in data transmitter periodically uploaded the data to a server. A computer running a scheduled preprocessing task picked up the data from the server, performed checks, corrections and calculations on the data (see section 2.2.1 SEMS Preprocessing) and wrote it in a dedicated SEMS database.

%
%In the first group are the sensors to measure \nox, \otwo and temperature and pressure sensors welded on the exhaust at the end of the tail-pipe together with a GPS receiver and antenna whereas in the second are data from the \cb such as velocity, throttle and rotational frequency %angular velocity
%of the engine. All the data acquired was gathered in a data logger located in the cabin as depicted in figure \ref{fig:truck}. Furthermore, this computer was set up to start (stop) acquisition when the ignition was turned on (off).
%
%Mass flow of both, \nox and \co are computed from the read of \nox and \otwo sensors described. In short, concentrations of \nox and \otwo are directly measured at the tailpipe. By means of the measurements from the mass air flow (MAF) intake from the \cb, the concentration of \otwo and \nox can be obtained. Finally, to calculate the concentration of \co and the fuel mass flow, the properties of the fuel and the ambient are used namely, the {C:H} ratio, the molar mass and the ambient oxygen concentration.
%All the data is stored at a frequency of $1\ \si{\hertz}$. Hence, the dataset consists of a number of aligned timeseries including speed, position and \co and \nox mass flow that start, and end based on the vehicle's ignition. 
%
\subsection{Data Processing}
\subsubsection{SEMS Preprocessing}
Data as collected from sensors generally need preprocessing to remove illogical values and spikes that cannot be explained by the conditions. Furthermore, some signals require calibration-based corrections. The variation in \cb signal implementation among manufacturers is another reason to perform checks and corrections.
The preprocessing of the SEMS data is performed automatically on a server, although some preemptive checks have been done already within the SEMS itself.
The following tasks are performed during preprocessing:
\begin{itemize}
\item Calibration; sensor data are corrected using a function based on calibration of the particular sensor in the lab.
\item Speed signals from GPS and vehicle are compared and combined to produce a good and continuous signal.
\item Raw cleaning; check for negative values, filter some signals for spikes, remove signals when the engine is not running, and fill small gaps in the data if possible.
\item Time alignment; align signals related to emissions to signals related to the engine.
\item Ammonia correction; correction of \nox concentration signal for ammonia, based on ammonia sensor values. This is necessary because the \nox sensor is cross sensitive for \nhthree.
\item Mass flow calculation; emissions in grams are calculated by calculating the flow of exhaust gas and multiplying it with the concentrations observed.
\end{itemize}
The result is a set of clean \SI{1}{\hertz} signals that can be used for further calculations, e.g. of emissions per km.

\subsubsection{Data Enrichment and Selection}
In order to enrich the data with information from the infrastructure, the time series of the GPS data needed to be connected to the road network. For each trip then the trajectory was \emph{map-matched} using the \emph{Open Source Routing Machine} (OSRM) \cite{luxen2011real}. The result of this is a new time series where each point is a node on \emph{Open Street Maps} (OSM) \cite{osm} from where any information present on it can be added to the original data.

\begin{figure}[H]
%\raggedleft
\centering
\includegraphics[width=.75\columnwidth]{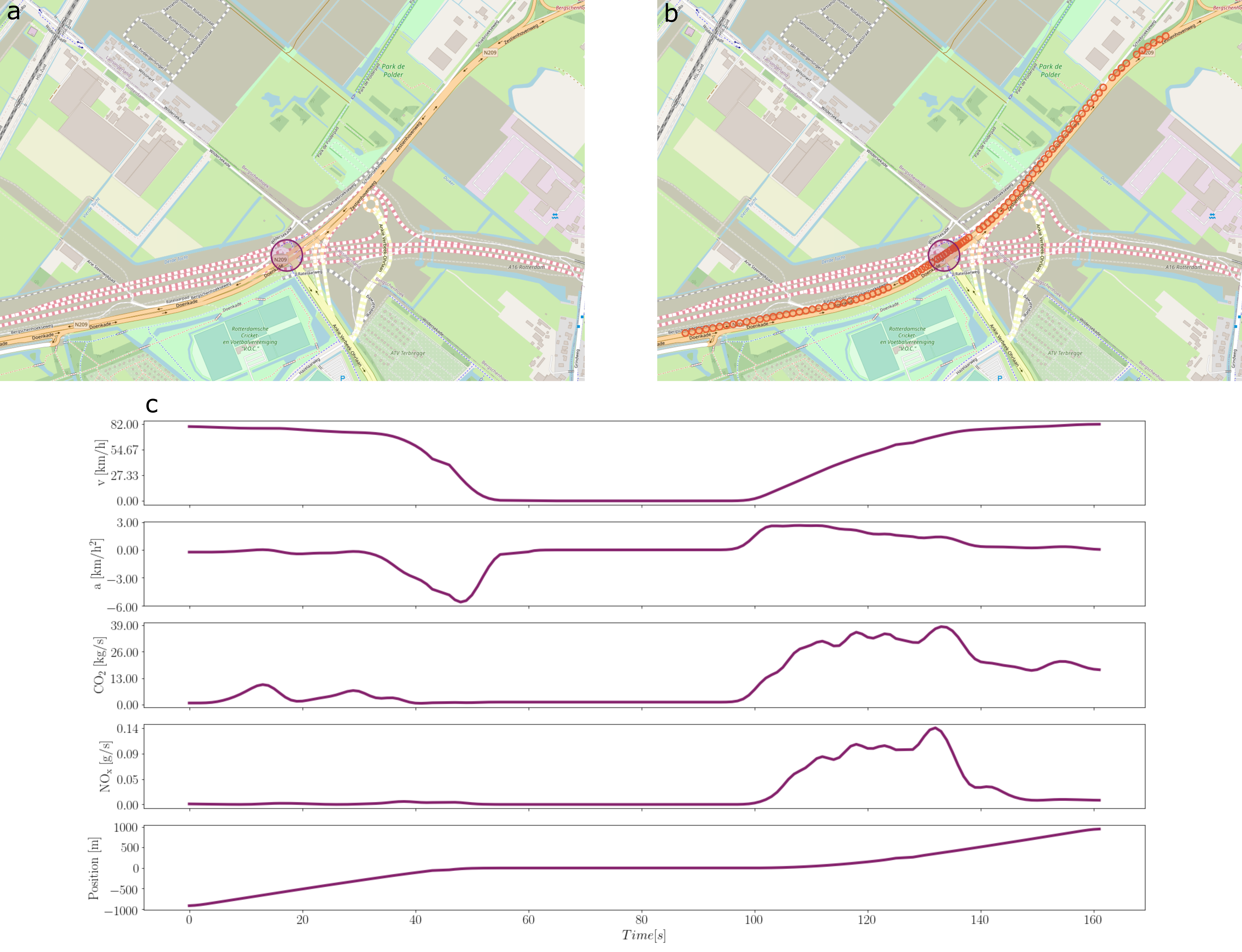}
\caption{Example of the extraction of the data around the intersections. (a) Signalized intersections where identified on the map. (b) The fragment of length \SI{2}{\km} symmetric around the intersection was selected based on the position data. (c) All the time series relevant for this study were gathered; from top to bottom, the velocity, acceleration, instantaneous \co and the instantaneous \nox. At the bottom, the position of the vehicle relative to the stop line is plotted for reference. There is a period between $60$ to \SI{100}{\second} in which the vehicle is waiting at the traffic light. Around \SI{100}{\second} is when the vehicle starts moving. Data spans only until the vehicle is \SI{1000}{\meter} past the intersection. }
\label{fig:segmentSelection}
\end{figure}

Once the data were linked to the map, we could identify all the points in the data in which a vehicle is crossing a signalized traffic light (based on traffic light locations as in OSM) and the respective segment of the road of equal length before and after the intersection. We identified the start and end points of segments of $\SI{2}{\km}$ length centered on the intersection crossing and also the vehicle action on the intersection i.e., the maneuver. Acknowledgedly arbitrarily, \SI{1000}{\meter} has been chosen based on the typical deceleration of a heavy-duty vehicle that can take up to \SI{1}{\km} \cite{maurya2012study,ligterink2016board}. From the original data the time series between the segment's start and end were retrieved; this process is schematized in Figure \ref{fig:segmentSelection}. The result is a series of $N$ intersection passages $p_i$ defining a set $P = \{p_1,\dots,p_N\}$. Each of these passages $d_i$ consists on group of time series of equal length representing the position ($x_i(t)$), the velocity ($v_i(t)$) and the instantaneous \co and \nox mass flow ($\xi_i(t)$ and $\theta_i(t)$), plus the maneuver taken by the vehicle (straight, right/left turn, u-turn) represented by $m_i$; i.e, $p_i = (x_i(t),v_i(t),\xi_i(t),\theta_i(t),m_i)$. The number of passages obtained was $N=11087$, from which only those with straight maneuver were picked, resulting in a smaller set $\tilde{P} = \{ p_i \in P \mid m_i= \mathrm{straight}\}$ with $| \tilde{P} | = 4972$. Provided the correlation between turn radius and vehicles velocity, the analysis of the turns is significantly more demanding and the interpretation of the results cumbersome; hence it was not carried out in this work and remains to be explored in future research.

%From the obtained new data set consisting of a series of 11063 intersection passages corresponding to $\SI{2}{\km}$ on the road, only those passages with a straight maneuver (4096) are used for the analysis. 

\subsection{Analysis}

\subsubsection{Clustering}
Three different clusters were defined from the speed profiles with a set of simple rules. These clusters represent different scenarios to be analyzed, a scenario in which the vehicle does not stop at the intersection and does not change its speed significantly; a scenario in which the vehicle has to stop completely at the intersection; and an intermediate scenario in which the vehicle decreases its speed significantly before the intersection but does not stop. Throughout the rest of the paper, we refer to these clusters as \textbf{no-stop}, \textbf{stop}, and \textbf{slow-down}; and we associate them with the colors green, red and amber, respectively. The clusters are based on the mean velocity of the vehicle in two main subsegments, the immediate last \SI{600}{\meter} before the intersection and the \SI{400}{\meter} that are between \SI{1000}{\meter} and \SI{600}{\meter} from the intersection. Defining for each passage $i$ the mean velocity of the subsegment determined by the interval $\psi = (a,b)$ as,
% We first define the mean velocities for these three subsegments as $\bar{v}_{-1000}^{-600}$, $v_{-600}^{0}$ and $v_{0}^{1000}$ where 
\begin{equation}
\bar{v}_i^{\psi} = \int_\psi \dif t v_i(t) = \int_a^b \dif t v_i(t),
\end{equation}
we have defined the mean velocities over five subsegments of interest. First, the three mean velocities $\bar{v}_i^\alpha$, $\bar{v}_i^\beta$ and $\bar{v}_i^\gamma$ for intervals $\alpha = [-600,-400)\, \si{\meter}$, $\beta = [-400,-200)\, \si{\meter}$ and $\gamma = [-200,50)\, \si{\meter}$, respectively are the three intervals used to classify the clusters. Two additional mean speeds $\bar{v}_i^\rho$, $\bar{v}_i^\omega$ for $\rho = [-1000,-600)\, \si{\meter}$, $\omega = [50,1000)\, \si{\meter}$ over the remaining parts of the passage were used to homogenize the behavior farther from the intersection. The same conditions are defined for the mean velocities $\bar{v}_i^\rho$, $\bar{v}_i^\omega$ for the three clusters. This reduces the general variability caused by external conditions such as traffic or weather. The three clusters are then defined based on a set of rules over $\bar{v}_i^\epsilon$ with $\epsilon \in \{\alpha,\beta,\gamma,\rho,\omega\}$ as follows,

%\begin{flalign}
%%\begin{equation}
% \begin{aligned}
% C_{\mathrm{no-stop}} &= \{ i \in P \mid &
% \\&&\bar{v}_i^\rho > 60\, \si{\kmh}
% \\&&\land\ \bar{v}_i^\epsilon > 60\, \si{\kmh}
% \forall \epsilon \in \{\alpha,\beta,\gamma\}
% \\&&\land\ \bar{v}_i^\omega > 30\, \si{\kmh}
% \\&&\land\ \nexists\ t \in [-1000,1000] \mid v_i(t) < 3\, \si{\kmh}
% \},\\
% C_{\mathrm{slow-down}} &= \{ i \in P \mid &
% \\&&\bar{v}_i^\rho > 60\, \si{\kmh}
% \\&&\land\ \bar{v}_i^\epsilon > 30\, \si{\kmh}
% \forall \epsilon \in \{\alpha,\beta,\gamma\}
% \\&&\land\ \exists \epsilon' \mid \epsilon' \in \{\alpha,\beta,\gamma\} 
% \land\ \bar{v}_i^\epsilon < \bar{v}_i^\rho\, \si{\kmh}
% \\&&\land\ \bar{v}_i^\omega > 30\, \si{\kmh}
% \\&&\land\ \nexists\ t \in [-1000,1000] \mid v_i(t) < 3\, \si{\kmh}
% \},\\
% C_{\mathrm{stop}} &= \{ i \in P \mid &
% \\&&\bar{v}_i^\rho > 60\, \si{\kmh}
% \\&&\land\ \bar{v}_i^\omega > 30\, \si{\kmh}
% \\&&\land\ \exists\ t \in \alpha \cup \beta \cup \gamma \mid v_i(t) < 3\, \si{\kmh}
% \\&&\land\ \nexists\ t \in \rho \cup \omega \mid v_i(t) < 3\, \si{\kmh}
% \},\\
% \end{aligned}
%%\end{equation}
%\label{eq:clusterDefinition}
%\end{flalign}

\begin{flalign}
%\begin{equation}
\begin{aligned}
C_{\mathrm{no-stop}} = \{ i \in P \mid&
\bar{v}_i^\rho > 60\, \si{\kmh}
\land\ \bar{v}_i^\epsilon > 60\, \si{\kmh}
\forall \epsilon \in \{\alpha,\beta,\gamma\}
\\&\land\ \bar{v}_i^\omega > 30\, \si{\kmh}
\land\ \nexists\ t \in [-1000,1000] \mid v_i(t) < 3\, \si{\kmh}
\},\\
C_{\mathrm{slow-down}} = \{ i \in P \mid &
\bar{v}_i^\rho > 60\, \si{\kmh}
\land\ \bar{v}_i^\epsilon > 30\, \si{\kmh}
\forall \epsilon \in \{\alpha,\beta,\gamma\}
\\&\land\ \exists \epsilon' \mid \epsilon' \in \{\alpha,\beta,\gamma\} 
\land\ \bar{v}_i^\epsilon < \bar{v}_i^\rho
\land\ \bar{v}_i^\omega > 30\, \si{\kmh}
\\&\land\ \nexists\ t \in [-1000,1000] \mid v_i(t) < 3\, \si{\kmh}
\},\\
C_{\mathrm{stop}} = \{ i \in P \mid &
\bar{v}_i^\rho > 60\, \si{\kmh}
\land\ \bar{v}_i^\omega > 30\, \si{\kmh}
\\&\land\ \exists\ t \in \alpha \cup \beta \cup \gamma \mid v_i(t) < 3\, \si{\kmh}
\land\ \nexists\ t \in \rho \cup \omega \mid v_i(t) < 3\, \si{\kmh}
\},\\
\end{aligned}
%\end{equation}
\label{eq:clusterDefinition}
\end{flalign}

\begin{figure}[H]
%\raggedleft
\centering
\includegraphics[width=.8\columnwidth]{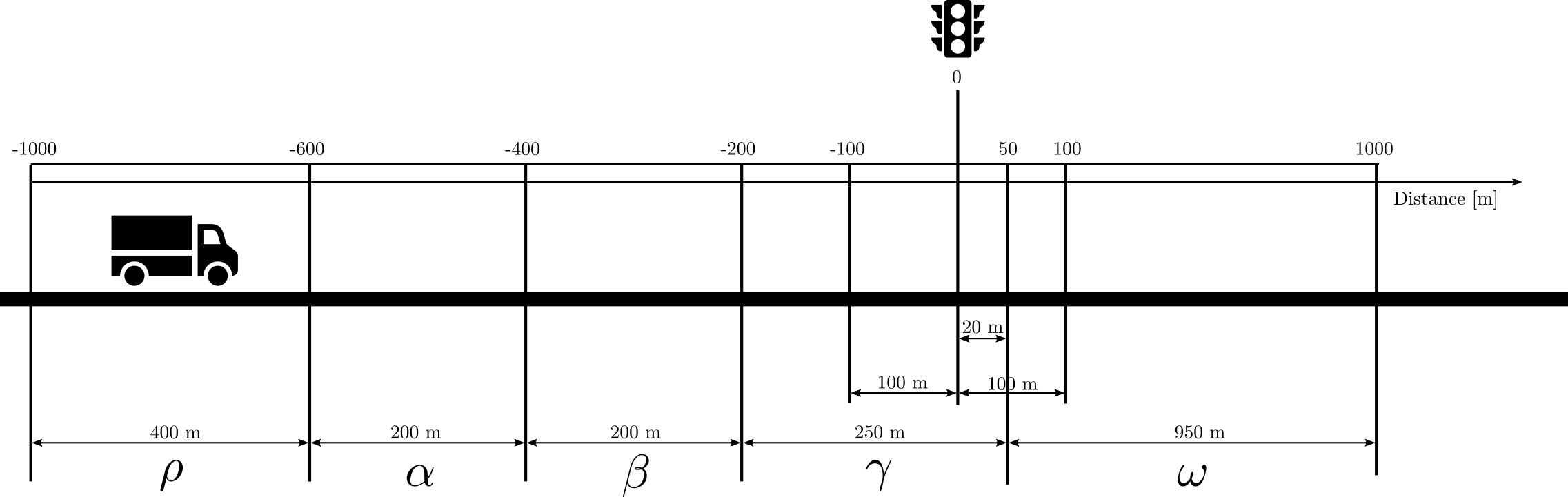}
\caption{Detail of the set of rules used in the study to determine the three clusters. The rules are mainly based on the approach to the intersection and they are significantly laxer on the conditions of the velocity after the intersection. In all cases, the vehicle mean velocity over the immediate \SI{400}{\meter} segment \SI{600}{\meter} away from the intersection is greater than \SI{60}{\km\per\hour}. The \SI{600}{\meter} prior to the intersection is where the clusters are distinguished; for the no-stop case, the mean velocity is greater than \SI{60}{\km\per\hour} over the whole \SI{600}{\meter} segment. For the stop cluster, the vehicle stops completely (the velocity vanishes) at least once in the segment. Finally, for the slow-down cluster, the mean velocity is lower than the mean velocity on the previous segment, but greater than \SI{30}{\km\per\hour} and furthermore the instantaneous velocity is greater than \SI{30}{\km\per\hour}. The mean velocity over the \SI{1000}{\meter} after the intersection for the vehicles in all the clusters need to be greater than \SI{30}{\km\per\hour}.}
\label{fig:clusterDefinition}
\end{figure}

These rules define the disjoint sets $C_{\mathrm{no-stop}}$, $C_{\mathrm{slow-down}}$, and $C_{\mathrm{stop}}$ that do not necessarily span the set of all the intersection passages, i.e., $C_{\mathrm{no-stop}} \cup C_{\mathrm{slow-down}} \cup C_{\mathrm{stop}} \subseteq C$. The results of the classification of the 4096 trajectories in $|\tilde{P}|$ are shown in Table \ref{tab:clusters}. A detailed graphical representation of the rules that defined the cluster is shown in Figure \ref{fig:clusterDefinition}.

\begin{table}[H]
\centering
\begin{tabular}{c||c}
Cluster & Number of passages \\ 
\hline
\hline
no-stop &378 \\ 
slow-down &349 \\ 
stop &175 \\
\end{tabular}
\caption{Number of passages classified in each of the three clusters after analyzing the $|\tilde{P}|=4096$ straight passages. It can be noted that the sum of the passages over the three clusters does not add to 4096 because not all the passages belong to a cluster.}
\label{tab:clusters} 
\end{table}

\subsection{Outcome Measures}

For all the intersection passages to be used in the analysis, the total \co and \nox over each \SI{2}{\km} segment was computed integrating the instantaneous quantities $\xi_i(t)$ and $\theta_i(t)$,
\begin{equation}
\Xi_i = \int_{s_i} \dif t \xi_i(t)\ \text{ for } i=1 \dots N,
\end{equation}
and 
\begin{equation}
\Theta_i = \int_{s_i} \dif t \theta_i(t)\ \text{ for } i=1 \dots N,
\end{equation}
where the lower case $\xi_i(t)$ and $\theta_i(t)$ are time-dependent quantities of passage $i$ and represent the instantaneous \co and \nox concentrations at the tail-pipe, respectively; $s_i$ is the total time interval that it takes the vehicle $i$ to transverse the entire \SI{2}{\km} passage; $N$ is the total number of passages included in the analysis; and the upper case $\Xi_i$ and $\Theta_i$ state the accumulated \co and \nox emitted over the entire intersection passage $i$.

Once $\Xi_i$ and $\Theta_i$ for all passages $i$ are computed, they are grouped by the clusters defined using the speed profiles (Equation \ref{eq:clusterDefinition}).
\begin{equation}
\begin{aligned}
\hat{\Xi}_{j} &= \{ \Xi_i | i \in C_\mathrm{j} \} \\
\hat{\Theta}_{j} &= \{ \Theta_i | i \in C_\mathrm{j} \},
% \hat{\xi}_{no-stop} &= \{ \Xi_i | i \in C_\mathrm{no-stop} \} \\
% \hat{\xi}_{slow-down} &= \{ \xi_i | i \in C_\mathrm{slow-down} \} \\
% \hat{\xi}_{stop} &= \{ \xi_i | i \in C_\mathrm{stop} \} \\
\end{aligned}
\end{equation}
for $j \in \{\mathrm{no-stop},\mathrm{slow-down},\mathrm{stop}\}$. These are the six sets we analyzed (two outcome measures for three groups). %\nameref{sec:StatisticalAnalysis}.
%\subsubsection{Statistical Analysis}
%\label{sec:StatisticalAnalysis}
Provided that the data are not normally distributed, the statistical significance of the results was assessed by means of a Kruskal-Wallis Rank test followed by a Pairwise Dunn test to determine the significance between the groups. Significance criterion was $p=0.01$.
%For each outcome the statistical significance of the results was assessed by means of an ANOVA test followed by a Pairwise Tukey test to determine the significance between groups. The data was tested for normality by means of a Shapiro-Wilk test.

%\section{Theory/calculation}
%\textcolor{red}{A Theory section should extend, not repeat, the background to the article already dealt with in the Introduction and lay the foundation for further work. In contrast, a Calculation section represents a practical development from a theoretical basis.}\\

\section{Results}
%\textcolor{red}{Results should be clear and concise.}\\
\subsection{Clustering of the speed profiles}
\begin{figure}[H]
%\raggedleft
\centering
\includegraphics[width=.5\columnwidth]{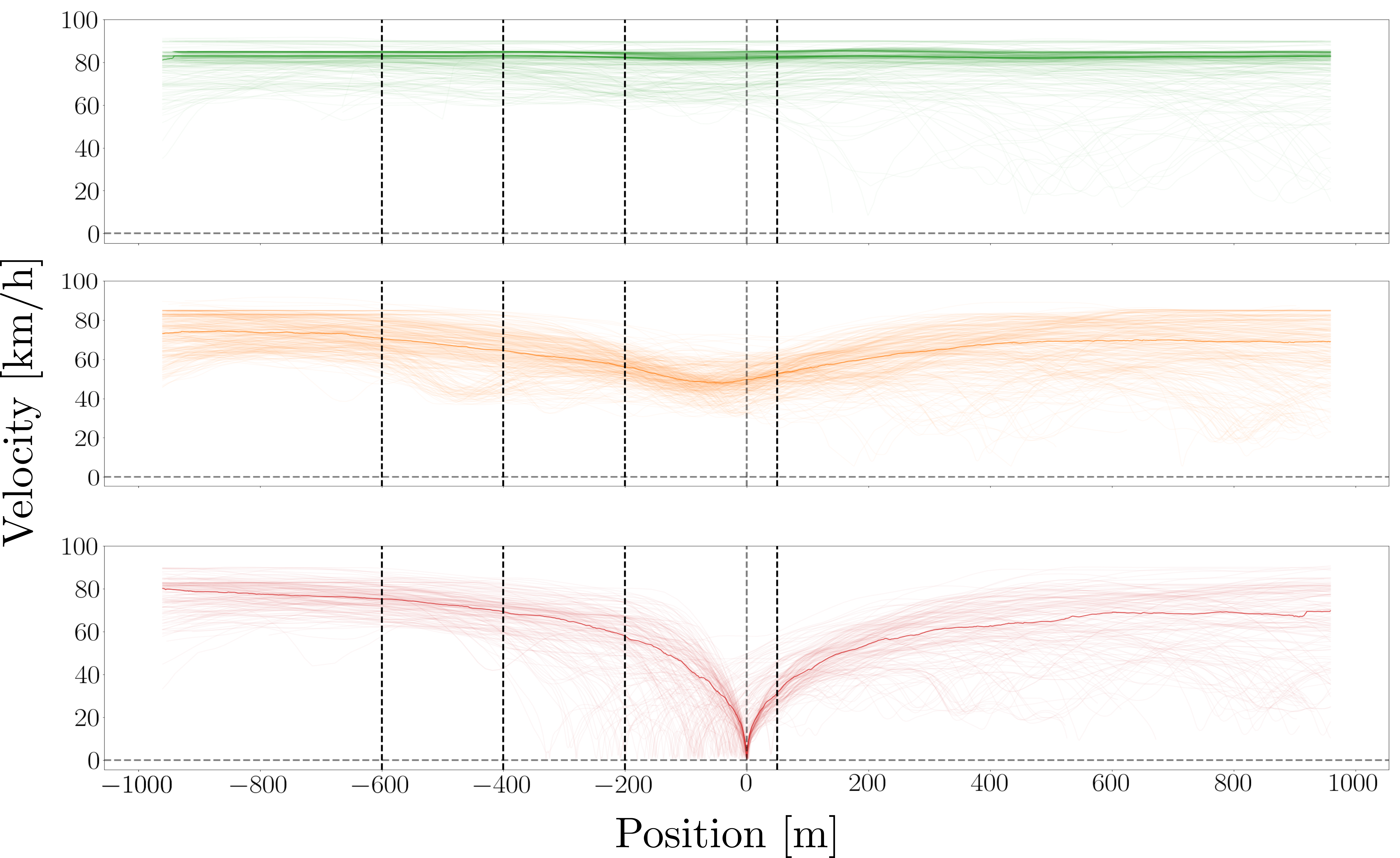}
\caption{Speed profiles for the three different groups. From top to bottom: stop, slow-down and no-stop clusters in red, amber and green, respectively. The speed profiles for all vehicles are included and shown in thinner lines. The thinner lines depict the median of each group.}
\label{fig:speedProfiles}
\end{figure}

The results of the clustering are shown in figure \ref{fig:speedProfiles}. The solid lines represent the median speed of each cluster along the \SI{\tslm}{\meter} segment whereas the shaded signifies the standard deviation. In green the no-stop cluster seem to keep a constant speed with relatively small deviation, while in amber, the slow-down cluster decreases its speed when approaching the intersection but does not stop. Finally, on the median of the stop cluster, it can be observed that the speed of the vehicles decreases significantly at the intersection. The reason why the median does not vanish is because not all the vehicles stop exactly at \SI{0}{\meter}, instead because of queues, many vehicles stop before it. It is also worth noticing that the standard deviations for all the clusters are smaller on the approach than after the intersection. This can be explained by the way the clusters are defined, where the rules apply to the approach and not to the segment after the intersection. The final speeds of the clusters then also differ, which we address in the Discussion section.

\subsection{\co emissions}
\begin{figure}[h!]
\centering
\includegraphics[width=\columnwidth]{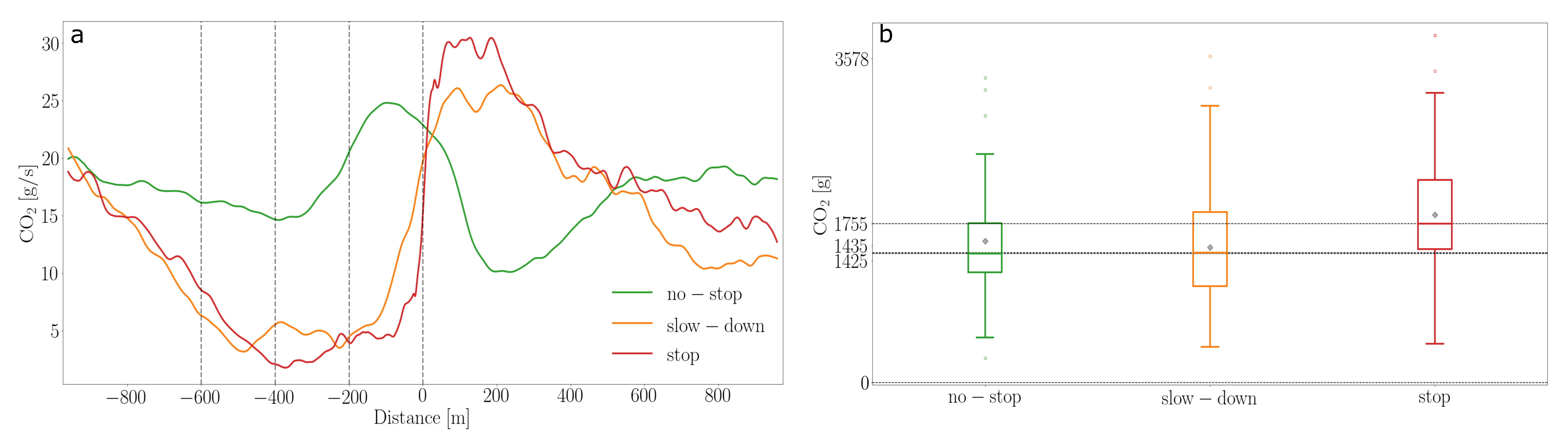}
\caption{\label{fig:co2}(a) Median instantaneous \co flow for the three groups. Red, amber and green for the stop, slow-down and no-stop groups, respectively. Position is relative to the location of the intersection as described on OSM. Vertical dashed lines denote the different regions used to group the speed profiles (see Methods\nameref{sec:methods} section). (b) Results for the \co flow for the three groups. Red, amber and green for the stop, slow-down and no-stop groups, respectively. Differences between all groups are significant.}
\end{figure}

Figure \ref{fig:co2}a. shows the instantaneous \co flow over the entire \SI{\tslkm}{\km} passage (medians over the entire cluster subpopulations). As expected, the slopes of the stop and slow-down clusters (in red and amber) right after the intersection are similar; in both cases, the vehicles accelerate full-throttle after the intersection. Since the vehicles in the slow-down cluster have a higher initial speed, the period of full throttle is shorter, and the peak of this cluster is lower. Finally, the green curve represents the no-stop cluster. Its median, although fairly more constant than those of the other two clusters, is not as flat as one could have expected, which is discuss further in the Discussion section. 
In Figure \ref{fig:co2}b. we present the median \co for each of the clusters over the entire \SI{2}{\km} passage. The largest differences emerge in the stop cluster compared to any of the other two clusters. The differences with the stop group are statistically significant for $p=0.01$. This indicates that the difference in \co emissions between the stop and no-stop is \SI{0.32}{\kg}, which translates into \SI{0.12}{\litre} of diesel fuel.
As can be seen in Figure \ref{fig:cum_both}a., when reaching the intersection, the \co levels emitted by the no-stop subpopulation are near constant, hence its cumulative levels linear, which at the intersection are higher than the other two which were coasting in this segment preceding the stop line. Furthermore, the stop cluster reaches the intersection with the lowest values and this enhances the impact of the stop and subsequent throttle: \SI{100}{m} after the intersection the cumulative \co values from the stop cluster are already higher than the other two.

%\begin{figure*}[h!]
%\centering
% \begin{subfigure}[t]{.45\columnwidth}
% \raggedleft
% \includegraphics[width=\columnwidth]{\imagePathTrue/meanFuelScenarios}
% \caption{}
%% \caption{Median instantaneous \co flow for the three groups. Red, amber and green for the stop, slow-down and no-stop groups, respectively. Position is relative to the location of the intersection as described on OSM. Vertical dashed lines denote the different regions used to group the speed profiles (see \nameref{sec:methods})}.
% \label{fig:instantco2}
% \end{subfigure}
% \begin{subfigure}[t]{.45\columnwidth}
% \raggedright
% \includegraphics[width=\columnwidth]{\imagePathTrue/fuelall}
% \caption{}
%% \caption{Results for the \co flow for the three groups. Red, amber and green for the stop, slow-down and no-stop groups, respectively. * indicates significant differences.}
% \label{fig:boxCo2}
% \end{subfigure}
% \caption{(a) Median instantaneous \co flow for the three groups. Red, amber and green for the stop, slow-down and no-stop groups, respectively. Position is relative to the location of the intersection as described on OSM. Vertical dashed lines denote the different regions used to group the speed profiles (see \nameref{sec:methods}). (b) Results for the \co flow for the three groups. Red, amber and green for the stop, slow-down and no-stop groups, respectively. * indicates significant differences.}
%\end{figure*}

\begin{figure}[h!]
\centering
\includegraphics[width=.5\columnwidth]{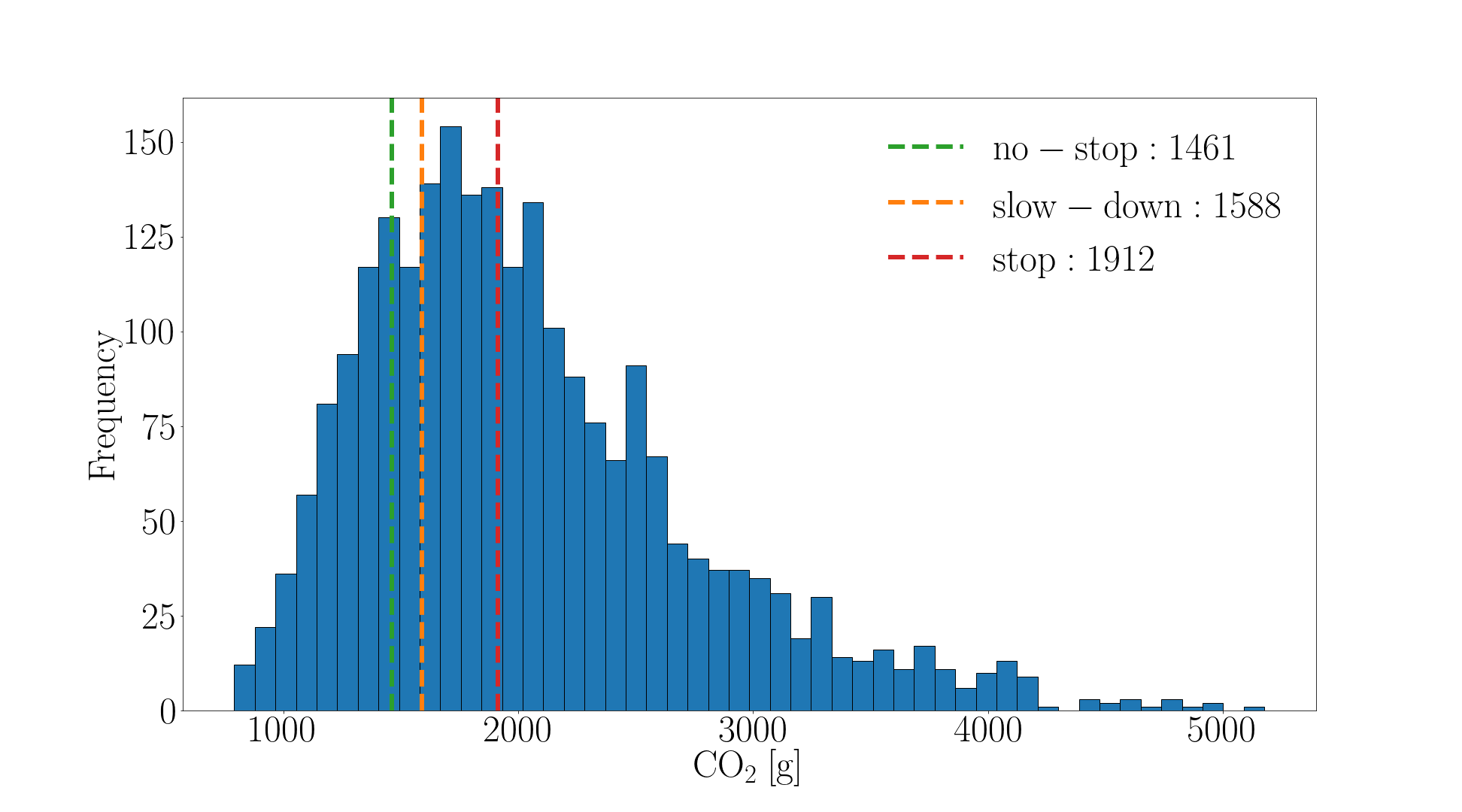}
\caption{\label{fig:distco2}Distribution of the \co emitted on the \SI{2}{\km} passage. In red, amber and green the median values for the no-stop, slow-down and stop clusters, respectively. The distribution is skewed and many values of emitted \co occur when interaction with other factors such as traffic is present. This indicates that the gap between the no-stopping vehicle and a vehicle that stops can increase.}
\end{figure}

\subsection{\nox emissions}
\begin{figure}[h!]
\centering
\includegraphics[width=\columnwidth]{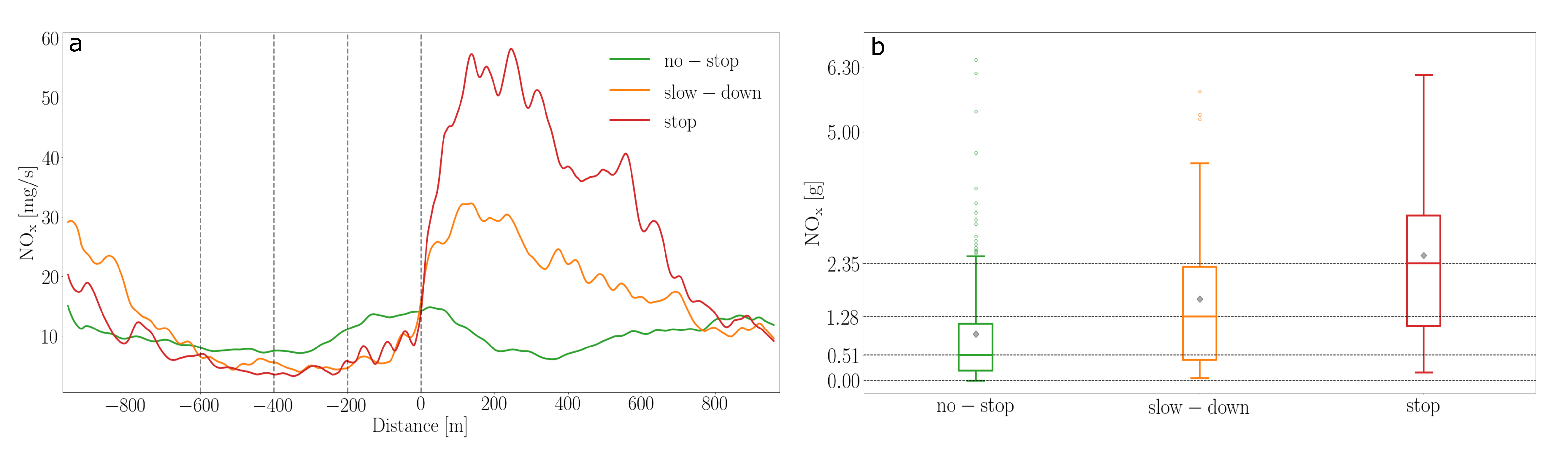}
\caption{\label{fig:nox}(a) Median instantaneous \nox flow for the three groups. Red, amber and green represent the stop, slow-down and no-stop groups, respectively. The position is relative to the location of the intersection as described on OSM. The vertical dashed lines denote the different regions used to group the speed profiles (see Methods section\nameref{sec:methods}). (b) Results for the \nox flow for the three groups. Red, amber and green represent the stop, slow-down and no-stop groups, respectively. Differences between all groups are significant.}
\end{figure}

\begin{figure*}[h!]
\centering
\includegraphics[width=\columnwidth]{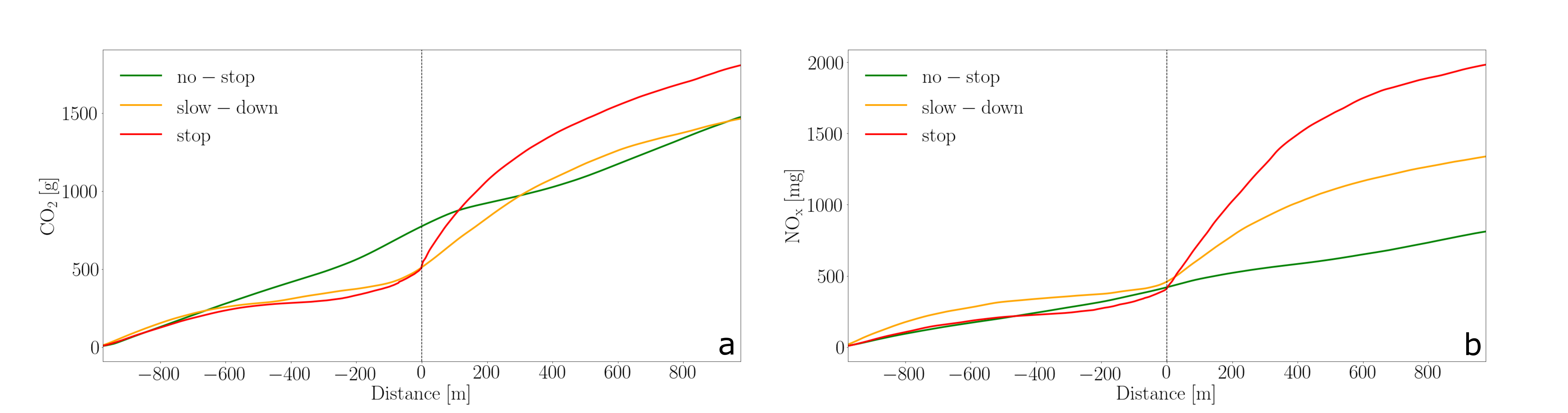}
\caption{\label{fig:cum_both}Cumulative flows for the three groups. Red, amber and green represent the stop, slow-down and no-stop groups, respectively. The position is relative to the location of the intersection as described in OSM. The vertical dashed line denotes the location of the intersection. (a) Results for the \co flow for the three groups; (b) Results for \nox for the three groups.}
\end{figure*}

The three clusters have very differentiated levels of \nox with a clear negative impact of the stop on the \nox levels (see Figure \ref{fig:nox}a.). As with the \co, the \nox levels are related to the acceleration and throttle pattern after the intersection, but contrary to that case, the \nox levels for the stop and slow-down and clearly distinguishable. As can be seen from the total median \nox emissions \ref{fig:nox} the stop population emits more than five times the same \nox in the \SI{2}{\km} passage than the slow down cluster. It is of main interest to contrast these values to the current regulations, in particular, those applying to the location of the measurements in the EU. As a reference, the current \nox emission limit for vehicles in the category Euro VI is \SI{0.46}{\g/\kWh}~\cite{eu-595-2009,eu-582-2011}, which for a typical HDV could be roughly translated into \SI[per-mode=symbol,sticky-per]{0.5}{\g\per\km} of \nox (assuming an average truck demand of \SI[per-mode=symbol,sticky-per]{1.1}{\kWh\per\km}).
%This indicates that, on average, when stopping and slowing down the vehicles would not comply with the current regulations. % \ND{ok, this is tricky because the 1.5 wont hold}
Figure \ref{fig:cum_both}b. shows a different story for the \nox compared to the \co. All changes in velocity are associated with an increase of the \nox levels, the cumulative \nox emitted by the no-stop cluster is never higher than that of the other two. Looking at the segment before the intersection, in which the slow-down cluster holds a lower speed, it can be interpreted that holding a lower constant speed is associated with higher \nox levels. Moreover, after the intersection, the longer full throttle from the stop cluster rises its cumulative value that overtakes the slow-down cluster \SI{200}{\m} after the intersection.

\section{Discussion}
%\textcolor{red}{This should explore the significance of the results of the work, not repeat them. A combined Results and Discussion section is often appropriate. Avoid extensive citations and discussion of published literature.}\\
The results indicate that avoiding a single stop in an HDV reduces \nox emissions by \SI{1.8}{\g} and \co emissions by \SI{0.32}{\kg}, saving \SI{0.12}{\litre} of diesel fuel. As explained in the Results section, this is true for the case of optimal traffic conditions, i.e., close to free flow with minimum interaction among vehicles where the vehicles approach and leave the intersection without being affected by traffic. 
%The results show that avoiding stops can save \SI{0.12}{\litre} of fuel at each intersection, without taking into account traffic, this means, these two clusters (stop and no-stop) are two cases in which the vehicle approach and leaves the intersection without being affect by traffic.
The cases analyzed are on the lower side of the distribution of emissions, see Figure \ref{fig:distco2}. This means that these figures represent a lower bound for the savings. When traffic plays a role, avoiding traffic congestion and the stop altogether would increase the savings considerably. %, up to \SI{}{\kg} for \nox and \SI{}{\kg} for \co.
In times of emerging C-ITS, this comes as a flexible and low-cost option for the governments to improve the sustainability of freight transport reducing emissions and costs. Programming the traffic lights adequately can make the difference between being \nox-emission compliant or not.

Due to the lack of homogeneity in the measuring methods, the different setups used and the arbitrary choice of the outcome measures, an accurate comparison of the results with other studies is not straightforward. Rough estimates can nevertheless be done for the studies that measured trucks in naturalistic behavior. The Compass4D project reported \co reductions in the range of \SIrange{20}{60}{\gram\per\km} \cite{compass4d2015}. This is in line with the results reported in the current work and it corresponds to the cases in which a vehicle avoids a stop every \SI{16}{\km} to \SI{5.3}{\km}. The Freilot project found aggregated levels of savings between 8\% and 13\% in measurements carried out over 12 months in France and The Netherlands, respectively. Based on the average distance travelled by a truck in 2012 as stated in Eurostat \cite{eurostat2012truckDistance} this translates into avoiding a stop every \SI{5}{\km} in The Netherlands and every \SI{8}{\km} in France; a result that is also in line with our current findings.

Naturally, several things remain to be improved and further explored. In our analysis, the clustering focused only on the approach to the intersection. This led to the deviation of the speed profiles after the intersection to be relatively high and furthermore, the final median speed for the different clusters is not the same. This introduces a bias in our results, making the difference between the clusters smaller. We nevertheless performed an analysis of this last part of the segments, and we found that the difference could not increase more than \SI{0.02}{\km} of \co or \SI{0.01}{\litre} of fuel (i.e., the \SI{0.32}{\km} and \SI{0.12}{\litre} difference of \co and fuel would be \SI{0.33}{\kg} and \SI{0.13}{\litre}) which does not affect the main message of our work.
Another important point to discuss is the increase and decrease of the fuel consumption before the intersection. We hypothesize that this increase in fuel consumption (preceded by a small decrease) is of behavioral nature. Drivers might release throttle when approaching until they are certain that they will reach the light in green and then, they accelerate again. This is particularly relevant because is a behavior that can be avoided with C-ITS technology that can guarantee priority to the drivers enabling them to keep a constant speed.
Finally, in the current analysis only the speed profiles were taken into account. A detailed analysis of the vehicle's complex dynamics while approaching an intersection, also taking into account driver’s behavior is still missing but needed for a full understanding. We encourage fellow researchers to explore that path.

% while giving priority to a vehicle the difference can only increase as we found when analyzing the \co emissions for all the passages that have been excluded in the clustering. This result becomes particularly important considering the government objectives and goals where reducing \co emissions while keeping logistics costs low is key on the shift toward the emissions goals at the same time that it does not create a negative impact on economy.

%On times of emerging C-ITS, this comes as a flexible and low-cost option for the governments to improve freight transport. 
\section{Conclusions}
The impact of signalized intersections on \nox and \co emissions in HDV was studied experimentally. It was concluded that avoiding a stop reduces the \nox emissions in a \SI{2}{\km} by $75 \%$. 
% the difference between not having to stop and stopping in a traffic light makes the \nox emissions go from below the EU limits to above them.
This is a key result for governments, regulatory authorities and traffic designers to consider. Furthermore, the \co levels can be reduced by \SI{0.32}{\km} per intersection. This represents more than $20 \%$ of the \co emitted at an intersection when stopping. This value represents a lower bound that would increase when including the interaction with traffic.
%\textcolor{red}{The main conclusions of the study may be presented in a short Conclusions section, which may stand alone or form a subsection of a Discussion or Results and Discussion section.}\\

%\section{Glossary}
%\textcolor{red}{Please supply, as a separate list, the definitions of field-specific terms used in your article.}\\

\section*{Funding}
%This study was partially funded by the \href{https://www.zuid-holland.nl/}{Province of Zuid Holland}and sponsored by the \href{https://mrdh.nl/}{MRDH} as part of the \href{http://catalystlab.nl/}{CATALYST project}.
This study was sponsored by the \href{https://mrdh.nl/}{MRDH} as part of the \href{http://catalystlab.nl/}{CATALYST project}.

% \section*{Acknowledgements}
% The authors want to thank the Catalyst program for providing the necessary environment to carry out the current research. ND wants to thank Jessica de Ruiter for help with the dataset, Paul R. Mentink for fruitful discussions and Norbert E. Ligterink for revising and objectively criticizing the manuscript. The authors want to thank Diederik Dijkema, Kin Fai Chan and Patrick van Norden for their feedback.

\bibliography{catalyst}
\bibliographystyle{plainnat}

\appendix
\section{Vehicle Mass Estimation}
\begin{figure*}[h!]
\centering
\includegraphics[width=.5\columnwidth]{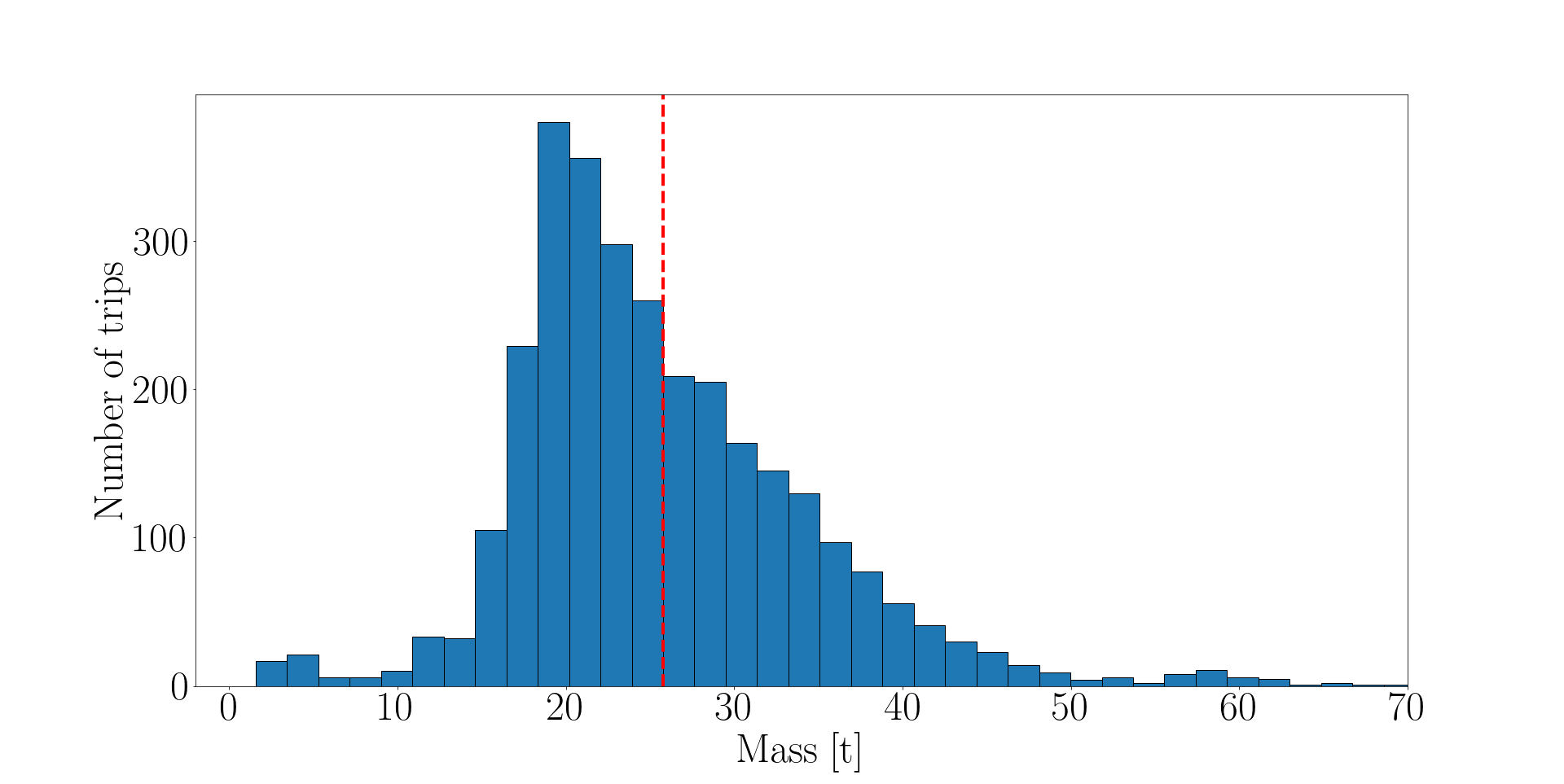}
\caption{\label{fig:massDist}Distribution of the total mass of the vehicles (tractor and load). The mean value is indicated in red.}
\end{figure*}

The total mass of the vehicles (tractor and load) was estimated to have a better description of the vehicles over which the analysis was done. The calculation used Newton's second law on the power,
\begin{equation}
P = m v * a + \alpha * m * v + \beta * v^2
\end{equation}
where $P$ is the power expressed in \si{\kW}, m the total mass in \si{\tonne}, v and a are the longitudinal velocity and acceleration in \si{\meter\per\second} and \si{\meter\per\second\squared}, respectively, and $\alpha$ and $\beta$ were parameters estimated on previous analysis to be $\alpha = 80 \si{\newton\per\tonne}$ and $\beta = 800 \si{\newton}$.
the power $P$ was computed from 
\begin{equation}
P = \frac{\tau 2 \pi \omega}{60}
\label{eq:power}
\end{equation}
using the measured engine rotational speed $\omega$ (in RPM) and the rated vehicle's maximum power, i.e., $\tau$ was set such that the maximum value of equation \ref{eq:power} matched the values on table \ref{tab:truckSpecs}. the mass distribution is shown in Figure \ref{fig:massDist}, the mean mass was \SI{31/3}{\tonne}.

\section{Statistical Analysis}
In order to evaluate the statistical significance of the differences of the means between the values of the different groups, several tests were done. First, it was evaluated whether the groups where normal distributed by means of a Shapiro-Wilk test, which for both \co and \nox happened not to be the case. Hence, non-parametric statistical tests were used namely, Kruskal-Wallis rank sum test \cite{kruskal1952use} and a Dunn pair test \cite{dunn1961multiple} as implemented in R \cite{rteam2017}.

Table \ref{tab:kw_co2} shows the results of the Kruskal-Wallis test for the \co. The p-value indicates that significant differences exist for a confidence level of $0.01$. Table \ref{tab:dunn_co2} shows the Dunn pairwise test for the \co. All groups show significant differences between each other for a significant value of $0.01$, and between the stop and the two other groups the differences are significant for more conservative significant levels.

Tables \ref{tab:kw_nox} and \ref{tab:dunn_nox} show the results of the same tests done on the \co but on the \nox values. Again, all groups show statistically significant differences for confidence level of $0.01$.

\begin{table}[h]
\begin{tabular}{ c |c| c }
$\chi^2$ &degrees of freedom &p-value \\
\hline
84.23 & 2& 2.2e-16 \\ 
\end{tabular}
\caption{Results of the Kruskal-Wallis test on the \co showing that significant differences exist.}
\label{tab:kw_co2}
\end{table}
\begin{table}[h]
\begin{tabular}{ c || c | c }
Scenario pair & mean rank difference & p-value \\ 
\hline
slow-down $\times$ no-stop & 52.64967 & 0.0037 \\ 
stop $\times$ no-stop & 197.80831 & 2e-16 \\ 
slow-down $\times$ stop & 145.15864 & 4.7e-10 \\ 
\end{tabular}
\caption{Results of the Dunn pairwise test on the \co. }
\label{tab:dunn_co2}
\end{table}

\begin{table}[h]
\begin{tabular}{ c |c|c }
$\chi^2$ & degrees of freedom & p-value \\ 
\hline
142.85 & 2 & 2.2e-16 \\ 
\end{tabular}
\caption{Results of the Kruskal-Wallis test on the \nox showing that significant differences exist.}
\label{tab:kw_nox}
\end{table}
\begin{table}[h]
\begin{tabular}{ c || c | c }
Scenario pair & mean rank difference & p-value \\ 
\hline
slow-down $\times$ no-stop & 141.68511 & 2e-16 \\ 
stop $\times$ no-stop & 203.46920 & 2e-16 \\ 
slow-down $\times$ stop & 61.78409 & 0.0019 \\ 
\end{tabular}
\caption{Results of the Dunn pairwise test on the \nox. }
\label{tab:dunn_nox}
\end{table}

%\textcolor{red}{If there is more than one appendix, they should be identified as A, B, etc. Formulae and equations in appendices should be given separate numbering: Eq. (A.1), Eq. (A.2), etc.; in a subsequent appendix, Eq. (B.1) and so on. Similarly, for tables and figures: Table A.1; Fig. A.1, etc.}\\

\end{document}